\documentclass[12pt,preprint]{elsarticle}
\usepackage[latin9]{inputenc}
\usepackage{amsmath}
\usepackage{amssymb}
\usepackage{esint}

\makeatletter
\@ifundefined{date}{}{\date{}}
\usepackage{amsfonts}\usepackage{color}
\usepackage{xcolor}\usepackage{graphicx}
\usepackage{subfigure}\usepackage{manfnt}
\usepackage{picinpar}

\usepackage{extarrows}
\usepackage{epstopdf}
\usepackage{picins}

\newcommand{\rom}[1]{\uppercase\expandafter{\romannumeral#1}}

\newcommand{\beq}{\begin{equation}}
\newcommand{\eeq}{\end{equation}}
\newcommand{\bal}{\begin{align}}
\newcommand{\eal}{\end{align}}
\newcommand{\baln}{\begin{align*}}
\newcommand{\ealn}{\end{align*}}

\graphicspath{{figdraft/}}
\journal{Journal of Computational Physics}

\makeatother

\begin{document}
\begin{frontmatter}

\title{Comment on ``Hamiltonian splitting for the Vlasov-Maxwell equations''}

\author[label1,label2]{Hong Qin\corref{cor1}}

\cortext[cor1]{hongqin@ustc.edu.cn}

\author[label1,label3]{Yang He}

\author[label1,label3]{Ruili Zhang}

\author[label1,label3]{Jian Liu}

\author[label1,label3]{ Jianyuan Xiao}

\author[label1,label3]{ Yulei Wang}

\address[label1]{ School of Nuclear Science and Technology and Department of Modern
Physics, University of Science and Technology of China, Hefei, Anhui
230026, China}

\address[label2]{Plasma Physics Laboratory, Princeton University, Princeton, NJ 08543,
USA}

\address[label3]{Key Laboratory of Geospace Environment, CAS, Hefei, Anhui 230026,
China}

\begin{abstract}
The paper \citep{Crouseilles15} by Crouseilles, Einkemmer, and Faou
used an incorrect Poisson bracket for the Vlasov-Maxwell equations.
If the correct Poisson bracket is used, the solution of one of the
subsystems cannot be computed exactly in general. As a result, one
cannot construct a symplectic scheme for the Vlasov-Maxwell equations
using the splitting Hamiltonian method proposed in Ref.\,\citep{Crouseilles15}. \end{abstract}
\begin{keyword}
splitting method \sep Vlasov-Maxwell system \sep symplectic integrator
\sep Poisson bracket 
\end{keyword}
\end{frontmatter}

In a recent paper \citep{Crouseilles15} by Crouseilles, Einkemmer,
and Faou, a new symplectic splitting method for the Vlasov-Maxwell
equations is proposed. In comparison with previous splitting methods,
the exciting new feature of the proposed method is that it is designed
to preserve the symplectic structure of the Vlasov-Maxwell system,
and thus enjoys the benefits of symplectic integration, such as the
global bound on energy error and long-term accuracy and fidelity. 

Crouseilles, Einkemmer, and Faou developed an innovative technique
to achieve this goal. A non-canonical Poisson bracket for the Vlasov-Maxwell
system as an infinite-dimensional Hamiltonian system is employed.
The system is split into three subsystems by splitting the Hamiltonian
functional into three parts. It turns out that the solution for each
subsystem can be computed exactly and therefore preserves exactly
the symplectic structure corresponding to the Poisson bracket. As
a consequence, the combined algorithm according to the splitting scheme
preserves the symplectic structure as well. In addition, higher order
methods can be constructed using various familiar composition methods.

The Poisson bracket adopted by Ref.\,\citep{Crouseilles15} is the
bracket discovered by Morrison in 1980 \citep{Morrison80}, 
\begin{align}
[F,G] & (E,B,f)=\int f\left\{ \frac{\delta F}{\delta f},\frac{\delta G}{\delta f}\right\} _{xv}dxdv\nonumber \\
 & +\int\left[\frac{\delta F}{\delta E}\cdot\left(\triangledown\times\frac{\delta G}{\delta B}\right)-\frac{\delta G}{\delta E}\cdot\left(\triangledown\times\frac{\delta F}{\delta B}\right)\right]dx\nonumber \\
 & +\int\left(\frac{\delta F}{\delta E}\cdot\frac{\partial f}{\partial v}\frac{\delta G}{\delta f}-\frac{\delta G}{\delta E}\cdot\frac{\partial f}{\partial v}\frac{\delta F}{\delta f}\right)dxdv\nonumber \\
 & +\int\left[\frac{\delta F}{\delta B}\cdot\left(\frac{\partial f}{\partial v}\times v\right)\frac{\delta G}{\delta f}-\frac{\delta G}{\delta B}\cdot\left(\frac{\partial f}{\partial v}\times v\right)\frac{\delta F}{\delta f}\right]dxdv,\label{eq:MB}
\end{align}
for functionals $F$ and $G$ of $E$, $B$, and $f.$ Here $\left\{ h,g\right\} _{xv}$
is the canonical Poisson bracket in the $(x,v)$ space. The Hamiltonian
for the system is 
\begin{equation}
H(f,E,B)=\frac{1}{2}\int v^{2}fdxdv+\frac{1}{2}\int\left(E^{2}+B^{2}\right)dx.
\end{equation}
This Hamiltonian can be split into three parts \citep{Crouseilles15}
as follows, 
\begin{alignat}{1}
H & =H_{f}+H_{E}+H_{B},\label{eq:Hs}\\
H_{f} & =\frac{1}{2}\int v^{2}fdxdv,\\
H_{E} & =\frac{1}{2}\int E^{2}dx,\\
H_{B} & =\frac{1}{2}\int B^{2}dx.
\end{alignat}
The scheme developed in Ref. \citep{Crouseilles15} is based on the
observation that solutions of the subsystems corresponding to $H_{f}$,
$H_{E}$, and $H_{B}$ can all be computed exactly. 

Unfortunately, this Poisson bracket \eqref{eq:MB} is known to be
incorrect, because it does not satisfy the Jacobi identity. This error
had been discovered and corrected \citep{Weinstein81,Marsden82} shortly
after its publication \citep{Morrison80}. The correct Poisson bracket
\citep{Weinstein81,Marsden82} is 
\begin{align}
[F,G] & (E,B,f)=\int f\left\{ \frac{\delta F}{\delta f},\frac{\delta G}{\delta f}\right\} dxdv\nonumber \\
 & +\int\left[\frac{\delta F}{\delta E}\cdot\left(\triangledown\times\frac{\delta G}{\delta B}\right)-\frac{\delta G}{\delta E}\cdot\left(\triangledown\times\frac{\delta F}{\delta B}\right)\right]dx\nonumber \\
 & +\int\left(\frac{\delta F}{\delta E}\cdot\frac{\partial f}{\partial v}\frac{\delta G}{\delta f}-\frac{\delta G}{\delta E}\cdot\frac{\partial f}{\partial v}\frac{\delta F}{\delta f}\right)dxdv\nonumber \\
 & +\int fB\left(\frac{\partial}{\partial v}\frac{\delta F}{\delta f}\times\frac{\partial}{\partial v}\frac{\delta G}{\delta f}\right)dxdv.\label{eq:MMWB}
\end{align}
Following Ref.\,\citep{Burby14-1970}, we will call this bracket
the Morrison-Marsden-Weinstein (MMW) bracket. Integrating the third
term on the right-hand side of Eq.\,\eqref{eq:MMWB} and considering
the fact that 
\begin{equation}
\frac{\partial}{\partial v}\left(\frac{\delta F}{\delta E}\right)=\frac{\partial}{\partial v}\left(\frac{\delta G}{\delta E}\right)=0,
\end{equation}
we can recast the MMW bracket as 
\begin{align}
[F,G] & (E,B,f)=\int f\left\{ \frac{\delta F}{\delta f},\frac{\delta G}{\delta f}\right\} dxdv\nonumber \\
 & +\int\left[\frac{\delta F}{\delta E}\cdot\left(\triangledown\times\frac{\delta G}{\delta B}\right)-\frac{\delta G}{\delta E}\cdot\left(\triangledown\times\frac{\delta F}{\delta B}\right)\right]dx\nonumber \\
 & +\int f\left(\frac{\delta G}{\delta E}\cdot\frac{\partial}{\partial v}\frac{\delta F}{\delta f}-\frac{\delta F}{\delta E}\cdot\frac{\partial}{\partial v}\frac{\delta G}{\delta f}\right)dxdv\nonumber \\
 & +\int fB\left(\frac{\partial}{\partial v}\frac{\delta F}{\delta f}\times\frac{\partial}{\partial v}\frac{\delta G}{\delta f}\right)dxdv.
\label{eq:MMWB2}
\end{align}
In Ref.\,\citep{Morrison81}, the MMW bracket in the form of Eq.\,\eqref{eq:MMWB}
is used, and in Refs.\,\citep{Morrison14,Chandre12}, the equivalent
form of Eq.\,\eqref{eq:MMWB2} is used. We emphasize that in order
for the MMW bracket in the form of Eq.\,\eqref{eq:MMWB} or Eq.\,\eqref{eq:MMWB2}
to satisfy the Jacobi identity, some constraints in terms of $B$
and/or $E$ are necessary. Marsden and Weinstein \citep{Marsden82}
restricted the solution space to be $Mv=\{(f,E,B)\mid\nabla\cdot B=0,\thinspace\nabla\cdot E=\int fdv\}.$
Morrison \citep{Morrison14} pointed out that it is only necessary
to require $\nabla\cdot B=0$ for the MMW bracket in the form of Eq.\,\eqref{eq:MMWB}
or Eq.\,\eqref{eq:MMWB2} to satisfy the Jacobi identity. Of course,
a solution of the Vlasov-Maxwell equations is always in $Mv$, if
it is initially in $Mv$.

If we use this correct Poisson bracket instead, in the form of either
Eq.\,\eqref{eq:MMWB} or Eq.\,\eqref{eq:MMWB2}, and apply the same
splitting scheme as in Eq.\,\eqref{eq:Hs}, it is very disappointing
to find out that the solution for the subsystem corresponding to $H_{f}$
can not be computed exactly, whereas the solutions for the subsystems
corresponding to $H_{E}$ and $H_{B}$ can. Specifically, the subsystem
associated with $H_{f}$ is 
\begin{alignat}{1}
\frac{\partial f}{\partial t} & +v\cdot\frac{\partial f}{\partial x}+\left(v\times B\right)\cdot\frac{\partial f}{\partial v}=0,\label{eq:fHf}\\
\frac{\partial E}{\partial t} & =-\int vfdv,\\
\frac{\partial B}{\partial t} & =0.\label{eq:fHB}
\end{alignat}
This subsystem is more complicated than its counterpart obtained using
the incorrect Poisson bracket \eqref{eq:MB}. Unless the magnetic
field $B$ is uniform in space or vanishes, its solution cannot be
computed exactly. As a result, for systems with general magnetic field,
one cannot construct a symplectic scheme for the Vlasov-Maxwell equations
using the splitting Hamiltonian method proposed in Ref.\,\citep{Crouseilles15}. 

If a symplectic integration method to a desired order for Eqs.\,\eqref{eq:fHf}-\eqref{eq:fHB}
can be found, then we can apply this splitting method to obtain a
symplectic scheme. However, such a symplectic method for Eqs.\,\eqref{eq:fHf}-\eqref{eq:fHB}
is not available yet. Further investigation is needed. 

As a final note, it is necessary to mention the following bracket
proposed by Chandre et al. \citep{Chandre12} to remove the $\nabla\cdot B=0$
constraint for the MMW bracket for the Vlasov-Maxwell equations,

\begin{align}
[F,G] & (E,B,f)=\int f\left\{ \frac{\delta F}{\delta f},\frac{\delta G}{\delta f}\right\} dxdv\nonumber \\
 & +\int\left[\frac{\delta F}{\delta E}\cdot\left(\triangledown\times\frac{\delta G}{\delta B}\right)-\frac{\delta G}{\delta E}\cdot\left(\triangledown\times\frac{\delta F}{\delta B}\right)\right]dx\nonumber \\
 & +\int f\left(\frac{\delta G}{\delta E}\cdot\frac{\partial}{\partial v}\frac{\delta F}{\delta f}-\frac{\delta F}{\delta E}\cdot\frac{\partial}{\partial v}\frac{\delta G}{\delta f}\right)dxdv\nonumber \\
 & +\int f(B-\nabla\triangle^{-1}\nabla\cdot B)\left(\frac{\partial}{\partial v}\frac{\delta F}{\delta f}\times\frac{\partial}{\partial v}\frac{\delta G}{\delta f}\right)dxdv.\label{eq:C}
\end{align}
Here, $B-\nabla\triangle^{-1}\nabla\cdot B$ is the projection of
$B$ that ``removes'' the non-divergence-free part of $B$. It is
straightforward to verify that the Jacobi identity is unconditionally
satisfied \citep{Chandre12}. For this bracket given by Eq.\,\eqref{eq:C},
the splitting Hamiltonian method proposed in Ref.\,\citep{Crouseilles15}
generates identical subsystems as for the MMW bracket given by Eq.\,\eqref{eq:MMWB}
or Eq.\,\eqref{eq:MMWB2}, when the $\nabla\cdot B=0$ constraint
is satisfied initially. Therefore, the splitting Hamiltonian method
proposed in Ref.\,\citep{Crouseilles15} is not valid for the bracket
given by Eq.\,\eqref{eq:C} as well.

\section*{Acknowledgments}

This research was supported by the National Natural Science Foundation
of China (11271357, 11261140328, 11305171), the CAS Program for Interdisciplinary
Collaboration Team, the ITER-China Program (2015GB111003, 2014GB124005,
2013GB111000), and the JSPS-NRF-NSFC A3 Foresight Program in the field
of Plasma Physics (NSFC-11261140328).


\end{document}